\documentclass[conference]{IEEEtran}
\IEEEoverridecommandlockouts

\usepackage{cite}
\usepackage{amsmath,amssymb,amsfonts}
\usepackage{algorithmic}
\usepackage[dvipdfmx]{graphicx}
\usepackage{textcomp}
\usepackage{url}
\usepackage{xcolor}

\ifCLASSOPTIONcompsoc
  \usepackage[caption=false,font=normalsize,labelfont=sf,textfont=sf]{subfig}
\else
  \usepackage[caption=false,font=footnotesize]{subfig}
\fi

\def\BibTeX{{\rm B\kern-.05em{\sc i\kern-.025em b}\kern-.08em
    T\kern-.1667em\lower.7ex\hbox{E}\kern-.125emX}}

\begin{document}

\title{Analysis of User Dwell Time on Non-News Pages}

\author{
    \IEEEauthorblockN{Ryosuke Homma\IEEEauthorrefmark{1} \quad Keiichi Soejima\IEEEauthorrefmark{2} \quad Mitsuo Yoshida\IEEEauthorrefmark{1} \quad Kyoji Umemura\IEEEauthorrefmark{1}}
    \IEEEauthorblockA{\IEEEauthorrefmark{1}\textit{Toyohashi University of Technology}\\
    Aichi, Japan\\
    r173360@edu.tut.ac.jp, yoshida@cs.tut.ac.jp, umemura@tut.jp}
    \IEEEauthorblockA{\IEEEauthorrefmark{2}\textit{Faber Company Inc.}\\
    Tokyo, Japan\\
    soejima@fabercompany.co.jp}
}

\maketitle

\begin{abstract}
There is dwell time as one of the indicators of user's behavior, and this indicates how long a user looked at a page.
Dwell time is especially useful in fields where user ratings are important, such as search engines, recommender systems, and advertisements are important.
Despite the importance of this index, however, its characteristics are not well known.
In this paper, we analyze the dwell times of various websites by desktop and mobile devices using data of one year.
Our aim is to clarify the characteristics of dwell time on non-news websites in order to discover which features are effective for predicting the dwell time.
In this analysis, we focus on device types, access times, behavior on the website, and scroll depth.
The results indicated that the number of sessions decreased as the dwell time increased, for both desktop and mobile devices.
We also found that hour and month greatly affected the dwell time, but day of the week had little effect.
Moreover, we discovered that inside and click users tended to have longer dwell times than outside and non-click users.
However, we can not find a relationship between dwell time and scroll depth.
This is because even if a user browsed the bottom of the page, the user might not necessarily have read the entire page.

\end{abstract}


\section{Introduction}
We need to know the browsing behavior of the user to improve web pages' quality.
There is dwell time as one of the indicators of user's behavior, and this indicates how long a user looked at a page.
Dwell time is used especially in fields where user ratings are important, such as search engines~\cite{Agichtein2006,Morita1994}, recommender systems~\cite{Yi2014}, and advertisements~\cite{Lalmas2015,Zhou2016} are important.
Despite the importance of this index, however, its characteristics are not well known.
Liu et al.~\cite{Liu2010} have analyzed the characteristics of dwell time, but they used data of only two weeks.
In addition, previous studies have reported that trends for news sites and applications vary greatly between desktop and mobile devices~\cite{Lalmas2015,Yi2014,Lagun2016},
but there has been little analysis focusing on non-news websites.

In this paper, we analyze the dwell times of various websites by desktop and mobile devices using data of one year.
Our aim is to clarify the characteristics of dwell time on non-news websites in order to discover which features are effective for predicting the dwell time.
To this end, we address the following research questions:
\begin{description}
 \item[RQ1] Dwell time by device:\\
   Is there a difference in the dwell time for each device?
 \item[RQ2] Dwell time by access time:\\
   Is there a difference in the dwell time depending hour of the day, day of the week, or month of the year?
 \item[RQ3] Dwell time by inside and outside:\\
   Is there a difference in the dwell time based on whether users visited from inside or from outside the site?
 \item[RQ4] Dwell time by click and non-click users:\\
   Is there a difference in the dwell time based on whether users clicked or did not click on the web pages?
 \item[RQ5] Dwell time by scroll depth:\\
   Is there a relationship between dwell time and scroll depth?
\end{description}

The results indicated that the number of sessions decreased as the dwell time increased, for both desktop and mobile devices.
The median of dwell time by desktop and mobile devices differed substantially when the websites were designed with only one type of device in mind.

We also found that hour and month greatly affected the dwell time, but day of the week had little effect.
In particular, a change in dwell time was observed based on the hour as well as the number of sessions (pageviews).

Moreover, we discovered that inside and click users tended to have longer dwell times than outside and non-click users.
The likely explanation for this result is that the user's amount of interest affects the length of dwell time.

However, we can not find a relationship between dwell time and scroll depth.
This is because even if a user browsed the bottom of the page, the user might not necessarily have read the entire page.

\section{Dataset}

\begin{table*}[tp]
  \caption{Websites used for analysis: The type, rate of the device, dwell time, and clicks are different for these websites.}
  \centering
  \begin{tabular}{c|l|l|rrr|rrr}
\hline
 & & & \multicolumn{3}{c|}{Desktop} & \multicolumn{3}{c}{Mobile} \\
\cline{4-9} 
\# & Data Period & Site Type & \multicolumn{1}{l}{Sessions} & \multicolumn{1}{l}{Time (s)} & \multicolumn{1}{l|}{Clicks} & \multicolumn{1}{l}{Sessions} & \multicolumn{1}{l}{Time (s)} & \multicolumn{1}{l}{Clicks} \\
\hline
    A & 2017/07/01 $\sim$ 2018/08/21 & Business & 117,329 (74.1\%) & 7 & 0.70 & 40,947 (25.9\%) & 8 & 0.63 \\
    B & 2017/07/02 $\sim$ 2018/08/21 & Blog & 10,291 (35.2\%) & 21 & 0.12 & 18,914 (64.8\%) & 23 & 0.13 \\
    C & 2018/04/11 $\sim$ 2018/08/21 & Blog & 192,879 (30.5\%) & 47 & 0.18 & 439,695 (69.5\%) & 72 & 0.18 \\
    D & 2017/08/02 $\sim$ 2017/08/31 & Affiliate & 724 (41.7\%) & 12 & 0.06 & 1,014 (58.3\%) & 12 & 0.00 \\
    E & 2017/07/01 $\sim$ 2018/08/02 & Business & 520,792 (82.3\%) & 23 & 0.53 & 112,126 (17.7\%) & 14 & 0.27 \\
    F & 2017/07/01 $\sim$ 2018/08/15 & Affiliate & 26,884 (38.8\%) & 7 & 0.09 & 42,425 (61.2\%) & 9 & 0.07 \\
    G & 2017/07/01 $\sim$ 2018/08/21 & Affiliate & 171,565 (25.6\%) & 55 & 0.27 & 498,439 (74.4\%) & 83 & 0.24 \\
    H & 2017/07/01 $\sim$ 2018/08/21 & BBS & 286,139 (44.6\%) & 7 & 0.77 & 355,166 (55.4\%) & 7 & 0.77 \\
    I & 2017/12/18 $\sim$ 2018/08/27 & Affiliate & 102,872 (40.6\%) & 26 & 0.04 & 150,399 (59.4\%) & 20 & 0.02 \\
    J & 2017/07/01 $\sim$ 2018/08/21 & Blog & 100,158 (70.2\%) & 37 & 0.16 & 42,474 (29.8\%) & 9 & 0.05\\
    K & 2017/07/01 $\sim$ 2018/08/21 & Affiliate & 51,340 (30.2\%) & 9 & 0.14 & 118,591 (69.8\%) & 35 & 0.18\\
    L & 2017/07/01 $\sim$ 2018/08/21 & Affiliate & 194,838 (27.5\%) & 10 & 0.05 & 514,863 (72.5\%) & 25 & 0.08\\
    M & 2017/07/01 $\sim$ 2018/04/27 & Affiliate & 18,073 (36.4\%) & 9 & 0.02 & 31,515 (63.6\%) & 10 & 0.02\\
\hline
  \end{tabular}
  \label{tab:dwell_time_inout}
\end{table*}

\subsection{Development}

In this study, we used the browsing data of 13 Japanese websites from July 2017 to August 2018.
This data was gathered using MIERUCA\footnote{\url{https://mieru-ca.com/en-us/}}, a web-marketing tool including the function to visualize user's behavior.
MIERUCA is a  well-known search engine optimization (SEO) support tools in Japan, provided by Faber Company Inc.\footnote{\url{https://www.fabercompany.co.jp/}}.

Our data consisted of four types: device, click, scroll, and read data.
The device data was recorded immediately after accessing the page, and included the device type (desktop, tablet, or mobile), the window size (width and height), the referrer URL, and the length of the page.
The length of the page was included in the device data because the length varied depending on the device used by the user.
The click data recorded the locations users clicked with the mouse cursor.
For mobile devices such as smartphones, the locations users tapped was recorded.
The scroll data indicated the location where the user stopped scrolling.
When the user stopped scrolling for more than 4 seconds, that location was recorded in the read data.
In addition, all data included the URL that the user accessed, the date of the action, and the session key.
The session key remains unchanged from the time a user accesses a page until leaving it.

Dwell time of each session was calculated as the difference between the first action date and the last action date.
In this study, we mainly analyzed the dwell time of each website.
We regarded the dwell time of a website as the median of the dwell times of all users who accessed the page.
A very long dwell time was not recorded, as it was assumed that the user left the web browser launched, 
Therefore, it was not appropriate to use the average value as a representative value of dwell time.
In addition, we calculated the dwell time by desktop devices (including tablets and laptops) and mobile devices (mainly smart phones) separately, in order to analyze the differences based on the device type.

\subsection{Basic statistics}

Basic website information is shown in TABLE~\ref{tab:dwell_time_inout}.
We used data from July 2017 to August 2018, but there were minor differences for each website.
Site type represents the type of website.
Due to the MIERUCA privacy policy limitations, we cannot list the actual website names.
Therefore, the websites are given anonymous letter identification (from A to M), and the type of website is included.
The types are affiliate, bulletin board system (BBS), blog, and business.
An affiliate type is a website mainly promoting the purchase of specific products and service subscriptions.
A BBS is a website to exchange messages between users.
Blogs provides information in a blog format.
Businesses offer specific functions and information for enterprises.
Sessions is the total number of sessions during the data period on the website,
time is the median of the dwell time of all sessions on the website,
and clicks is the mean of the number of clicks for all sessions for that website.
These values were calculated for both desktop and mobile devices separately.

The rate of the device differed for each website, but on the whole, mobile devices tended to have a longer dwell time.
However, desktop devices tended to have more clicks.
There was no correlation between the number of sessions (page views), the dwell time, and the clicks.

\section{Results and Discussion}

\subsection{RQ1: Dwell time by device}

Song et al.~\cite{Song2013} have reported that the dwell time in a search engine (\textit{i.e.}, Bing\footnote{\url{https://www.bing.com/}}) varies widely depending on the device used.
We analyzed whether the dwell time in non-news pages varied depending on the device used.
In addition, if differences between devices were characteristic of dwell time, we analyzed what kind of settings the website utilized.

\begin{figure}[tp]
  \centering
  \includegraphics[width=0.99\linewidth]{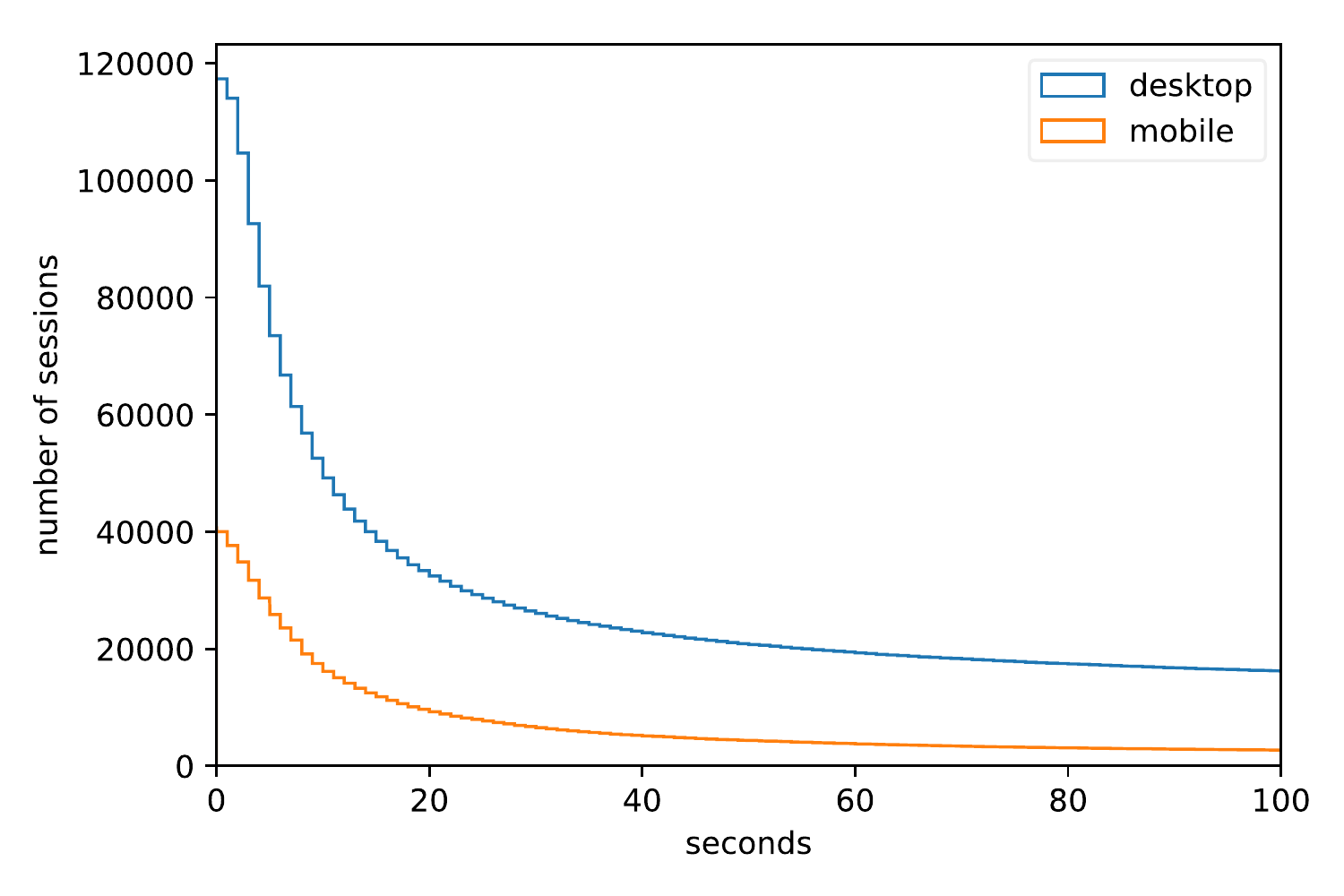}
  \caption{Site A: Dwell time and the number of sessions by device. As the dwell time increased, the number of sessions decreases, for both desktop and mobile devices.}
  \label{fig:number_of_users_dwell_time}
\end{figure}

The relationship between the number of sessions (users) and the dwell time for Website A is shown in Fig.~\ref{fig:number_of_users_dwell_time}.
In the previous study, Liu et al.~\cite{Liu2010} have shown that the relationship between the number of sessions and the dwell time for desktop devices followed the Weibull distribution.
In this study, the number of sessions decreased as the dwell time increased, for both desktop and mobile devices.
We plan to test whether these followed the Weibull distribution or not in a future study.

Depending on the website, the dwell time by device could differ greatly.
In many cases, a particular website was designed with only one type of device in mind.
For example, on Website J, the dwell time for desktop devices was 37 seconds, but for mobile devices is 9 seconds.
In other words, users with desktop devices stayed on pages of website J for much longer.
Since this site did not set a viewport with html meta tag, it seems that the design did not consider mobile users.
The viewport is a setting value for controlling the window size of the page.
Conversely, on Website C, the dwell time for the desktop devices is 47 seconds, while for mobile devices it was 72 seconds.
Since this site set a viewport with html meta tag, it seems that the design considered mobile users.

\subsection{RQ2: Dwell time by access time}

We analyzed whether the dwell time varied depending on the access time.
For access time, we focused on the hour of the day, day of the week, and the month of the year.
It is known that the number of accesses (page views) varies depending on the access time,
but it was not known whether the dwell time would change based on access time.

\begin{figure}[tp]
  \centering
  \subfloat[Site G: Example of website which dwell time changes.]{
    \includegraphics[width=0.99\linewidth]{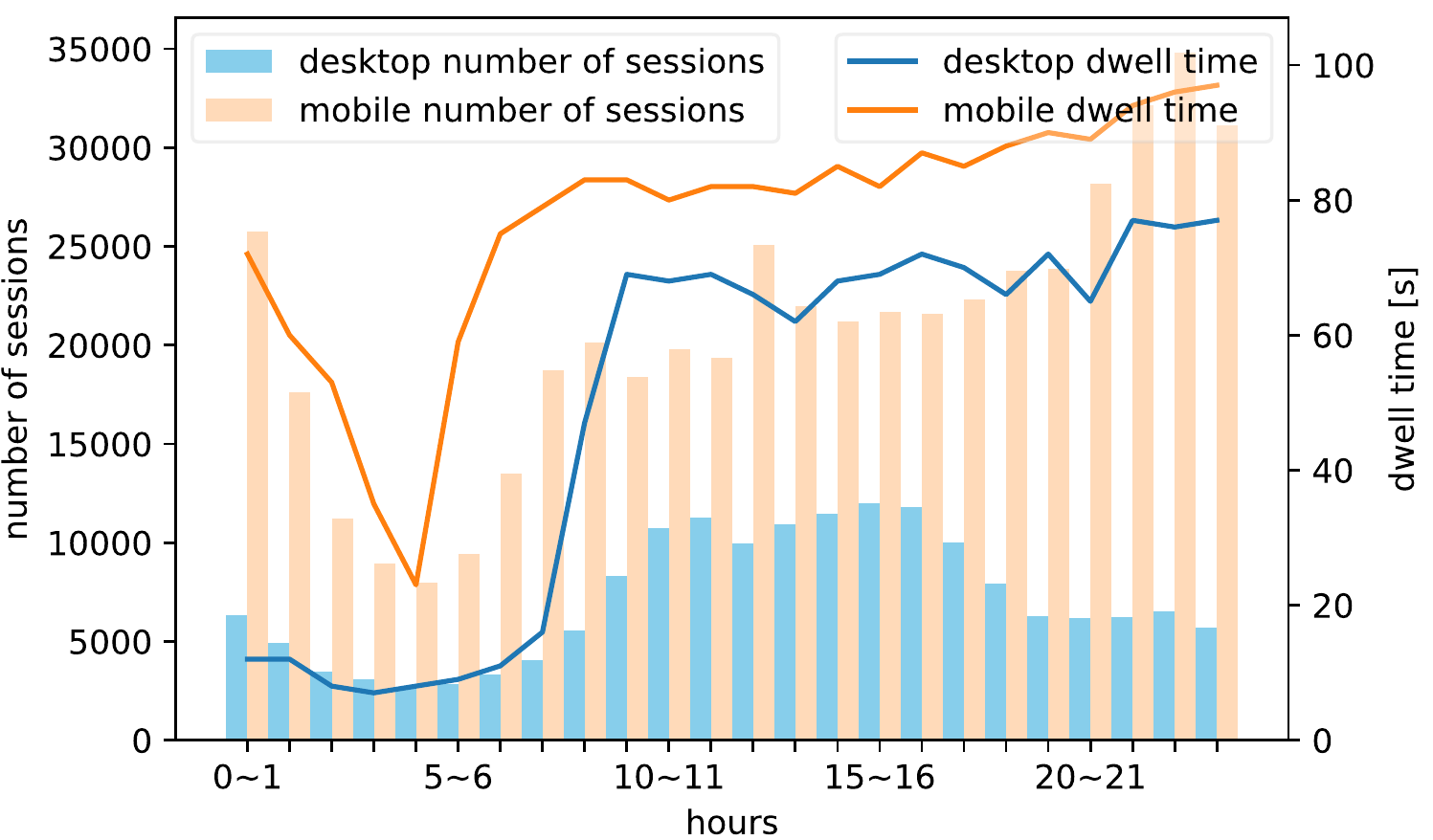}
    \label{fig:dwell_time_every_hour_G}
  }
  \hfil
  \subfloat[Site F: Example of website which dwell time does not change.]{
    \includegraphics[width=0.99\linewidth]{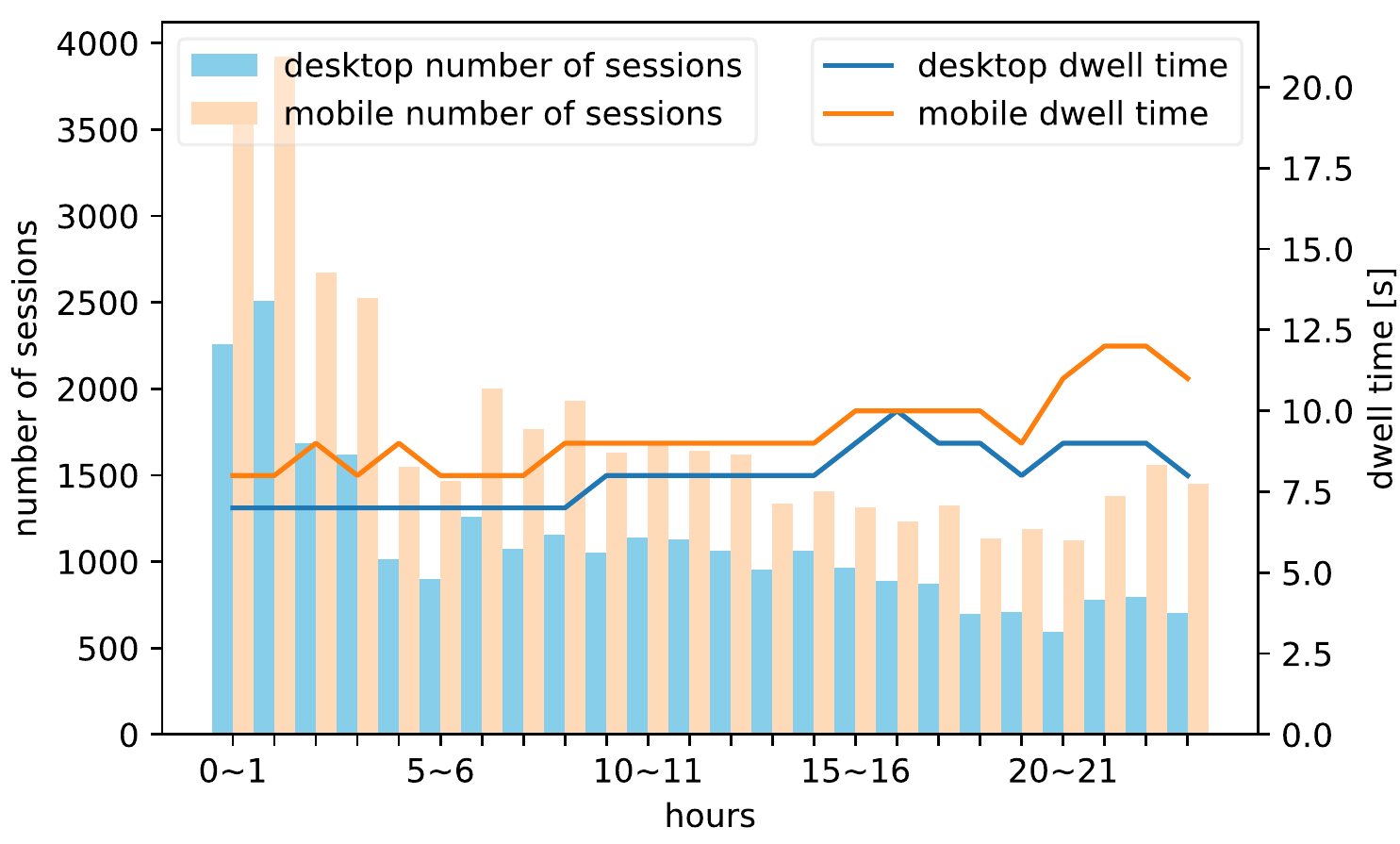}
    \label{fig:dwell_time_every_hour_F}
  }
  \caption{Dwell time by hour. The number of sessions tended to change by hour. For example, dwell time was low in the early morning and longer at night. For about half of the websites, the dwell time changed by hour.}
  \label{fig:dwell_time_every_hour}
\end{figure}

The dwell times by hour on Websites F and G are shown in Fig.~\ref{fig:dwell_time_every_hour}.
In these figures, the number of sessions is shown in addition to the dwell time.
Fig.~\ref{fig:dwell_time_every_hour_G} is an example where the dwell time changed, and Fig.~\ref{fig:dwell_time_every_hour_F} is an example where it did not change.
The number of sessions tended to change by hour; for example, there were less sessions in the early morning and more at night.
For about half of the websites, the dwell time changed by hour.
Blogs tended to have large change in dwell time, while business websites did not show large changes.

\begin{figure}[tp]
  \centering
  \subfloat[Site I: Example of website which dwell time changes.]{
    \includegraphics[width=0.99\linewidth]{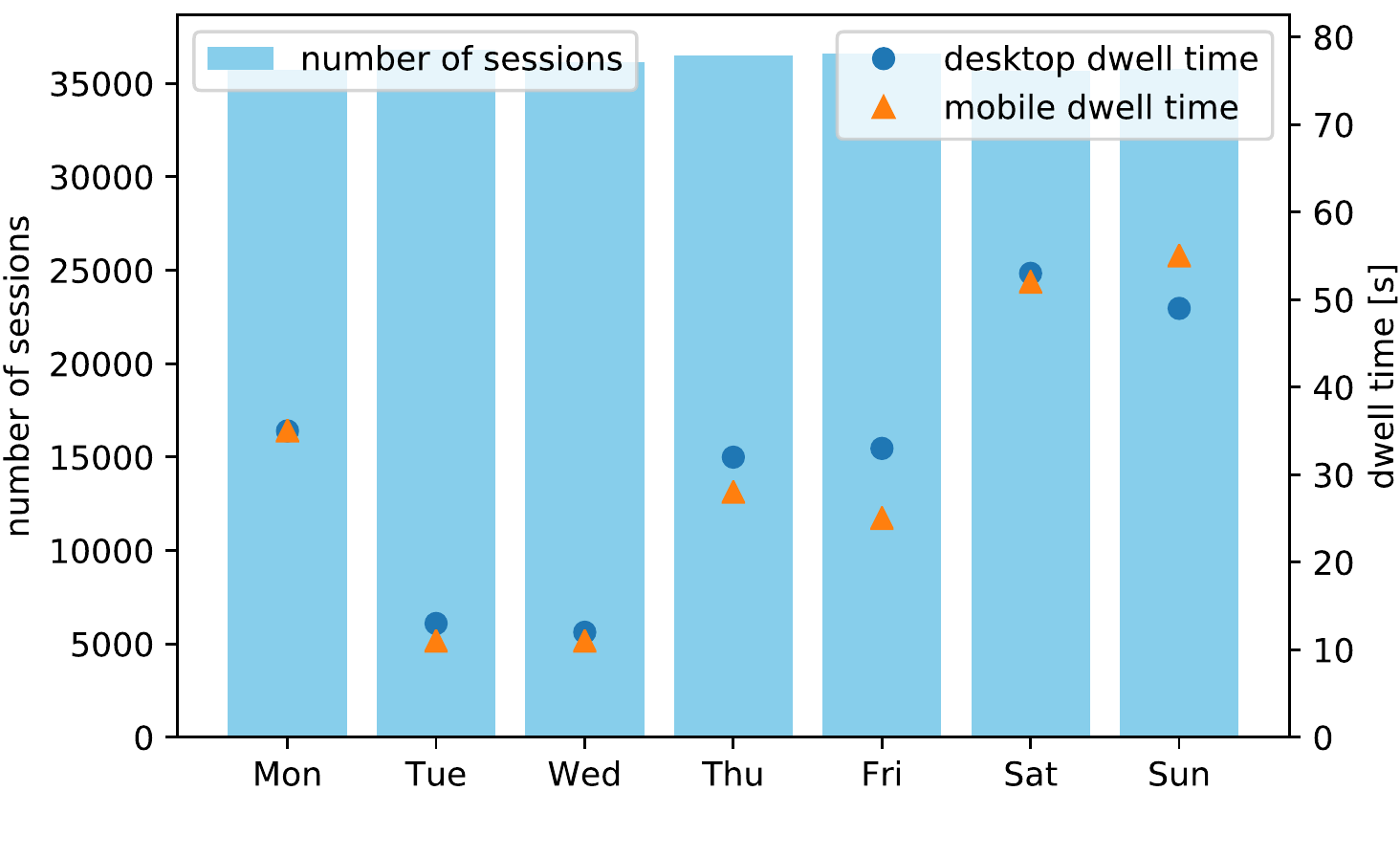}
    \label{fig:dwell_time_every_weekday_I}
  }
  \hfil
  \subfloat[Site B: Example of website which dwell time does not change.]{
    \includegraphics[width=0.99\linewidth]{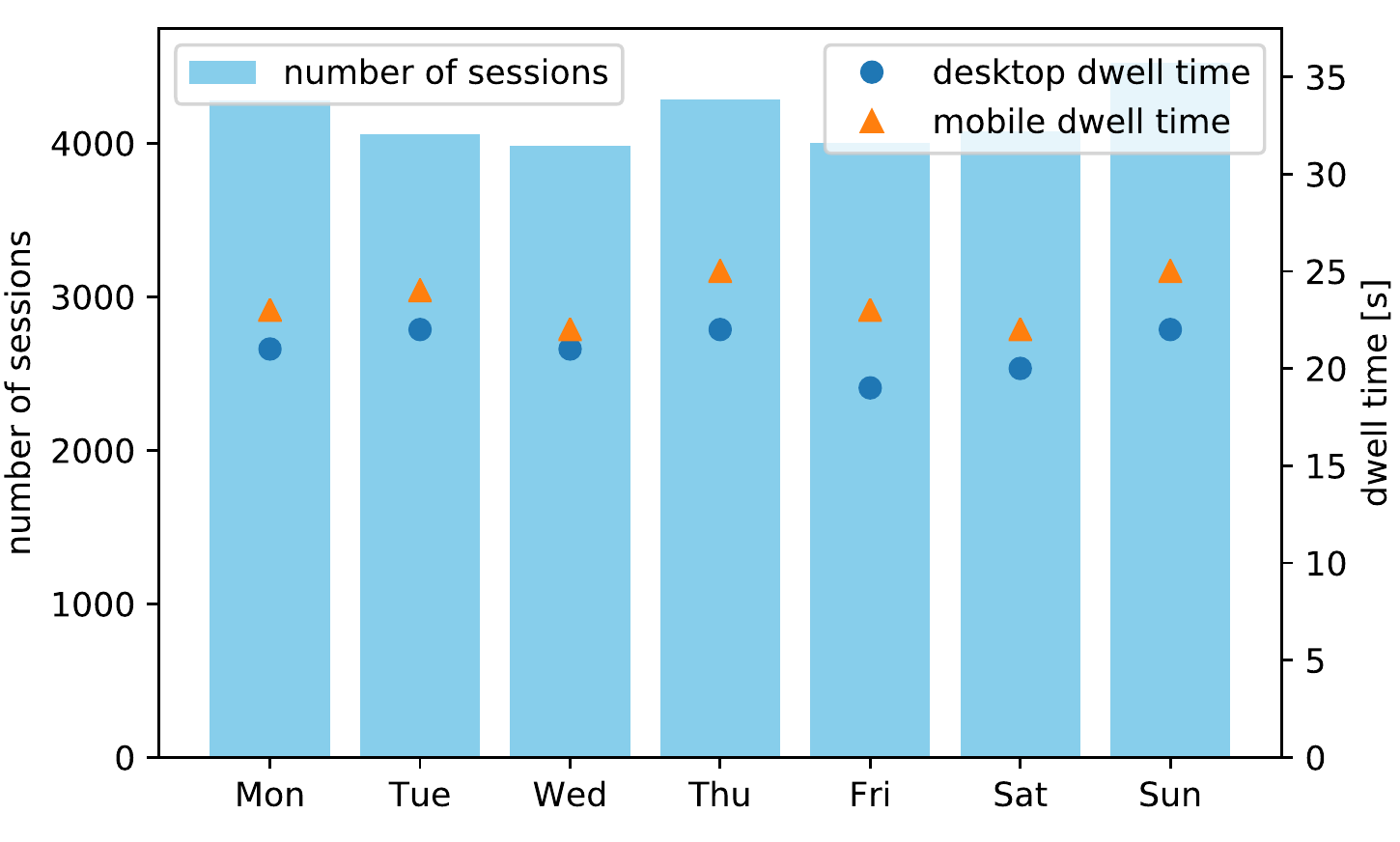}
    \label{fig:dwell_time_every_weekday_B}
  }
  \caption{Dwell time by day of week. The only website for which the dwell time changed greatly was Site I, which gives price information for used cars. The browsing behavior of the user on this site is affected by their intention to purchase a car.}
  \label{fig:dwell_time_every_weekday}
\end{figure}

The dwell times by day of week on Websites B and I are shown in Fig.~\ref{fig:dwell_time_every_weekday}.
In this figure, the number of sessions is shown in addition to the dwell time.
Fig.~\ref{fig:dwell_time_every_weekday_I} is an example where the dwell time changed, and Fig.~\ref{fig:dwell_time_every_weekday_B} is an example where it did not change.
The only website for which the dwell time changed greatly was Website I.
This website promotes price information for used cars.
When the user was preparing to take action (\textit{e.g.}, to visit a sales stores),
it could have affected the browsing behavior of the user, thus increasing the dwell time.

\begin{figure}[tp]
  \centering
  \subfloat[Site G: Example of website which dwell time changes.]{
    \includegraphics[width=0.99\linewidth]{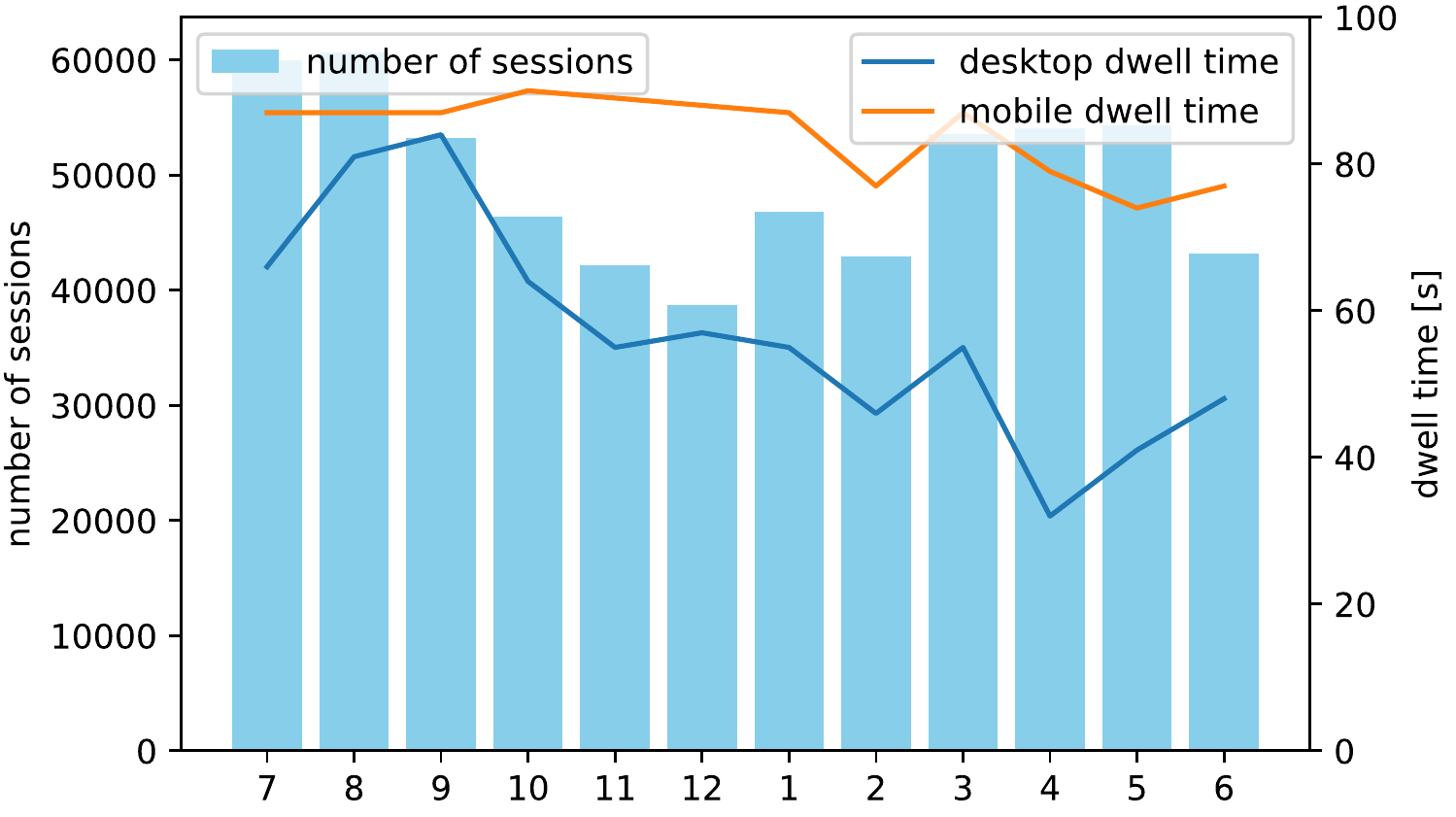}
    \label{fig:dwell_time_every_month_G}
  }
  \hfil
  \subfloat[Site A: Example of website which dwell time does not change.]{
    \includegraphics[width=0.99\linewidth]{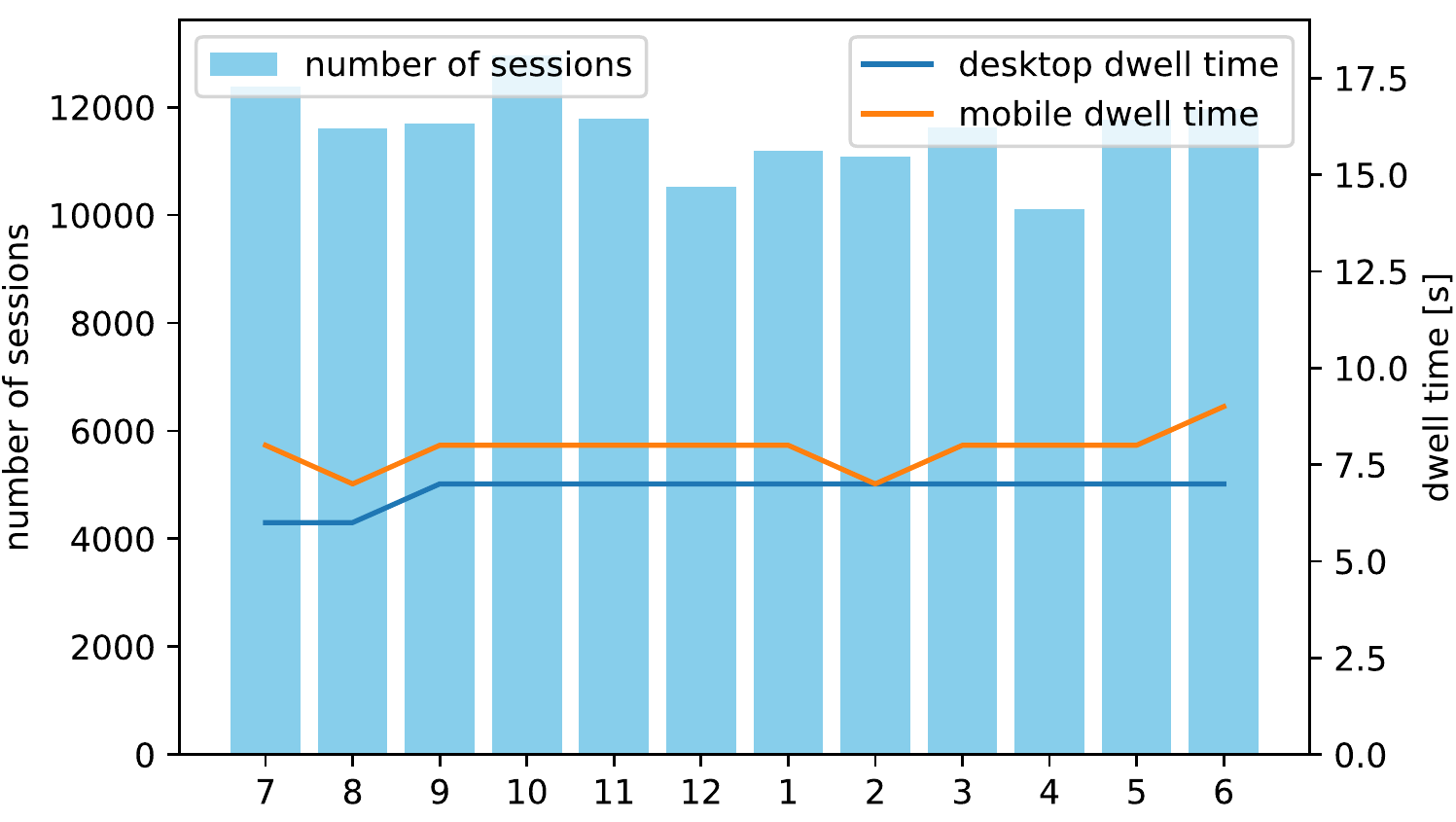}
    \label{fig:dwell_time_every_month_A}
  }
  \caption{Dwell time by month. Although there are both websites which change the dwell time by month and websites which do not change, we can not find those characteristics.}
  \label{fig:dwell_time_every_month}
\end{figure}

The dwell times by month on Website A and G are shown in Fig.~\ref{fig:dwell_time_every_month}.
In these figures, the number of sessions is shown in addition to the dwell time.
Fig.~\ref{fig:dwell_time_every_month_G} is an example where the dwell time changed, and Fig.~\ref{fig:dwell_time_every_month_A} is an example where it did not change.
we can not discover a reason for the differences in results between the websites.
It can be hypothesized that if the website is updated or popular content is added to the website, it will change the dwell time.
To test this, we will need to analyze the contents of websites in a future study.

A summary of the change in the dwell time by access time for each website is shown in TABLE~\ref{tab:dwell_time_access_time}.
In this table, Y indicates there was a change of 20 seconds or more for that field.
If there was a difference of more than 20 seconds between the dwell time at 1 o'clock and the dwell time at 8 o'clock,
we inserted Y in the hour column.
Site D had data for only one month, so there was no monthly change.
The BBS and business websites tended to not show changes in the dwell time by hour or day of the week.
For these sites, although the number of users (sessions) changed depending on hour and day of the week,
users' purpose for visiting websites did not changed, and thus the dwell time did not change.

\begin{table}[tp]
  \caption{Summary of change in dwell time by access time: In this table, Y indicates that there was a change of 20 seconds or more for that field. Site D only had data for one month, so there was no monthly change.}
  \centering
  \begin{tabular}{c|ccc}
\hline
    \# & Hour & Day & Month \\
\hline \hline
    A &   &   & \\ \hline
    B & Y &   & Y  \\ \hline
    C & Y &   &    \\ \hline
    D &   &   & -  \\ \hline
    E &   &   & Y  \\ \hline
    F &   &   &    \\ \hline
    G & Y &   & Y  \\ \hline
    H &   &   &    \\ \hline
    I & Y & Y & Y  \\ \hline
    J & Y &   &    \\ \hline
    K & Y &   & Y  \\ \hline
    L & Y &   & Y  \\ \hline
    M & Y &   &    \\ \hline
  \end{tabular}
  \label{tab:dwell_time_access_time}
\end{table}

\subsection{RQ3: Dwell time by inside and outside}

We analyzed whether the dwell time varied depending on if users visited from inside or from outside the site.
For this paper, we defined inside users as those who visited from the same website as the analysis target page.
Outside users were defined as those who visited from another website.

The dwell times of inside and outside users for each website is shown in TABLE~\ref{tab:dwell_time_inoutside}.
Inside users tended to have longer dwell times than outside users.
Users browsing multiple pages on a website are likely more interested in that website than others.
The level of interest affects the length of dwell time, and it is assumed that the dwell time of inside users is longer than outside users.
There tended to be less inside users than outside users.
Only BBS websites had more inside users than outside users.
The reason for this could be that the users stay in these websites to communicate with other users. 

\begin{table*}[tp]
  \caption{Dwell time by inside and outside the site: Inside users tend to have longer dwell time than outside users.}
  \centering
  \begin{tabular}{c|rr|rr|rr|rr}
\hline
 & \multicolumn{4}{c|}{Desktop} & \multicolumn{4}{c}{Mobile} \\
\cline{2-9}
 & \multicolumn{2}{c|}{Inside User} & \multicolumn{2}{c|}{Outside User} & \multicolumn{2}{c|}{Inside user} & \multicolumn{2}{c}{Outside user} \\ 
\# & \multicolumn{1}{l}{Sessions} & \multicolumn{1}{l|}{Time (s)} & \multicolumn{1}{l}{Sessions} & \multicolumn{1}{l|}{Time (s)} & \multicolumn{1}{l}{Sessions} & \multicolumn{1}{l|}{Time (s)} & \multicolumn{1}{l}{Sessions} & \multicolumn{1}{l}{Time (s)} \\
\hline
A & 14,283 &7 &103,046 &7 &8,159 &7 &32,788 &8 \\
B & 694&121.5 &9,597 &20 &400 &86 &18,514 &23 \\
C & 11,935&65 &180,944 &46 &7,656 &101 &432,039 &71 \\
D & 8& 86.5& 716& 12& 1& 14& 1,013& 12\\
E & 113,371& 40& 407,421& 20& 11,668& 13& 100,458& 14\\
F & 2,148& 40& 24,736& 7& 3,062& 39& 39,363& 9\\
G & 30,314& 79& 141,251& 50& 123,728& 56& 374,711& 90\\
H & 164,223& 7& 121,916& 7& 202,771& 7& 152,395& 6\\
I & 1,757& 18& 101,115& 28& 1,376& 17& 149,023& 20 \\
J & 8,887& 30& 91,271& 38& 1,210& 24& 41,264& 8\\
K & 6,496& 47& 44,844& 8& 15,374& 51& 103,217& 32\\
L & 20,895& 82& 173,943& 9& 28,830& 73& 486,033& 21\\
M & 193& 88& 17,880& 9& 412& 69.5& 31,103& 10.0\\
\hline
  \end{tabular}
  \label{tab:dwell_time_inoutside}
\end{table*}

\subsection{RQ4: Dwell time by click and non-click}

We analyzed whether the dwell time varied depending on if users clicked on the web pages or not.
In this paper, we defined click users as those who clicked on the analysis target page, whereas those who did not click on the analysis target page were defined as non-click users.
We did not consider the number of clicks for each user.

The dwell times of click and non-click users for each website are shown in TABLE~\ref{tab:dwell_time_click}.
Click users tended to have longer dwell times than non-click users.
When a user takes special action, in this case the action is clicking on a page, it means that they are likely interested in this page.
The level of interest affects the length of dwell time, and it is assumed that the dwell time of click users was longer.
There tended to be less click users than non-click users, with similar numbers for both desktop and mobile devices.
On the other hand, in only three websites (A, D, and H), click users had shorter dwell time than non-click users.
Site A and H tended to have pages with many links.
Users who routinely access such pages know the link destination and may be moving (clicking) in a short time.
Site D had only 17 click users in this study, and we will need more click users for a more detail analysis.

\begin{table*}[tp]
  \caption{Dwell time by click and non-click users: Click users tend to have longer dwell time than non-click users.}
  \centering
  \begin{tabular}{c|rr|rr|rr|rr}
\hline
 & \multicolumn{4}{c|}{Desktop} & \multicolumn{4}{c}{Mobile} \\
\cline{2-9}
 & \multicolumn{2}{c|}{Click User} & \multicolumn{2}{c|}{Non-Click User} & \multicolumn{2}{c|}{Click User} & \multicolumn{2}{c}{Non-Click User} \\ 
\# & \multicolumn{1}{l}{Sessions} & \multicolumn{1}{l|}{Time (s)} & \multicolumn{1}{l}{Sessions} & \multicolumn{1}{l|}{Time (s)} & \multicolumn{1}{l}{Sessions} & \multicolumn{1}{l|}{Time (s)} & \multicolumn{1}{l}{Sessions} & \multicolumn{1}{l}{Time (s)} \\
\hline
A & 68,292 &6 & 49,037&10 &22,164 &8 &18,783 &8 \\
B & 999& 136& 9,292& 19& 1,905& 83& 17,009& 21\\
C & 27,922& 123& 164,957& 34& 63,439& 122& 376,256& 62\\
D & 17& 5& 707& 12& 0& -& 1,014& 12\\
E & 132,655& 43& 388,137& 18& 11,974& 74& 100,152& 14\\
F & 2,021& 56& 24,863& 7& 2,755& 56& 39,670& 9\\
G & 34,747& 94& 136,818& 44& 91,994& 135& 406,445& 71\\
H & 216,383& 7& 69,756& 10& 267,790& 6& 87,376& 10\\
I & 2,965& 62& 99,907& 17& 2,478& 63& 147,921& 14\\
J & 12,661& 55& 87,497& 34& 1,849& 47& 40,625& 8\\
K & 5,639& 70& 45,701& 8& 17,930& 95& 100,661& 20\\
L & 8,304& 46& 186,534& 10& 35,839& 131& 479,024& 19\\
M & 199& 93& 17,874& 9& 446& 190& 31,069& 10\\
\hline
  \end{tabular}
  \label{tab:dwell_time_click}
\end{table*}

For click users, the dwell time until clicking and dwell time since clicking each website are shown in TABLE~\ref{tab:dwell_time_before_after_click}.
Dwell time since clicking tended to be shorter than dwell time until clicking.
If the user clicked the link of the page to go to another page, then the dwell time since clicking was zero.
Such users reduced the median of dwell time since clicking.
In other words, users going to another page within the same website are interested in this page, and thus were browsing for a long period of time.

\begin{table}[tp]
  \caption{Dwell time of click users: Dwell time since clicking tends to be shorter than dwell time until clicking.}
  \centering
  \begin{tabular}{c|rr|rr}
\hline
 & \multicolumn{2}{c|}{Desktop} & \multicolumn{2}{c}{Mobile} \\
\cline{2-5}
\# & \multicolumn{1}{l}{Until (s)} & \multicolumn{1}{l|}{Since (s)} & \multicolumn{1}{l}{Until (s)} & \multicolumn{1}{l}{Since (s)} \\
\hline
A & 5& 0&7 &0  \\
B & 49& 17& 21& 35\\
C & 23& 38& 20& 58\\
D & 4& 1& -& -\\
E & 20& 1& 32& 9\\
F & 46& 1& 48& 1\\
G & 73& 0& 95& 0\\
H & 6& 0& 6& 0\\
I & 24& 2& 18.5& 12\\
J & 40& 1& 38& 12\\
K & 52& 1& 75& 1\\
L & 21& 1& 55& 6\\
M & 52& 1& 128& 10.5\\
\hline
  \end{tabular}
  \label{tab:dwell_time_before_after_click}
\end{table}

\subsection{RQ5: Dwell time by scroll depth}

We analyzed whether the dwell time varied depending on how depth user's scroll depth on the web pages.
In this paper, we defined users who scrolled between 0\% and 25\% of the page as the top users, those who scrolled between 26\% and 75\% of the page as the middle users, and those who scrolled between 76\% and 100\% as the bottom users.
Users were classified into the top, middle, or bottom users based on the maximum depth for that session.

The dwell times of top, middle, and bottom users for each website are shown in TABLE~\ref{tab:dwell_time_how_deep}.
The top user's dwell times tended to be the shortest.
For about half of the websites, the bottom user's dwell times were the longest.
Thus, some users would scroll to the bottom of the page but have a low dwell time.
Even if a user browsed to the bottom of the page, the user may not necessarily have read the entire page.

\begin{table}[tp]
  \caption{Dwell time of scroll users: The top user's dwell time tends to be the shortest. However, it is about half of the websites where the bottom user's dwell time is the longest.}
  \centering
  \begin{tabular}{c|rrr|rrr}
\hline
 & \multicolumn{3}{c|}{Desktop} & \multicolumn{3}{c}{Mobile} \\
\cline{2-7}
\# & \multicolumn{1}{l}{Top} & \multicolumn{1}{l}{Middle} & \multicolumn{1}{l|}{Bottom} & \multicolumn{1}{l}{Top} & \multicolumn{1}{l}{Middle} & \multicolumn{1}{l}{Bottom} \\
\hline
A & 6& 25& 44& 7& 30& 47\\
B & 24& 26& 20& 26& 76& 15\\
C & 12& 42& 89& 16& 116& 163\\
D & 6.5& 18& 12& 304.5& 19& 11\\
E & 11& 48& 86& 14& 15& 14\\
F & 7& 8& 7& 8& 9& 9\\
G & 13& 56& 87& 11& 78& 131\\
H & 5& 11& 22& 6& 16& 21\\
I & 67& 24& 9& 67& 9& 20\\
J & 27& 43& 35& 8& 9& 7\\
K & 7& 7& 10& 9& 40& 54\\
L & 11& 9& 10& 18& 61& 16\\
M & 8& 10& 9& 9& 23& 9\\
\hline
  \end{tabular}
  \label{tab:dwell_time_how_deep}
\end{table}

\section{Conclusion}

We analyzed the dwell times of various websites by desktop and mobile devices using data of one year.
Our aim is to clarify the characteristics of dwell time on non-news websites in order to discover which features are effective for predicting dwell time.
We focused on device types, access times, behaviors within the website, and scroll depth in this analysis.

The results indicated that the number of sessions decreased as the dwell time increased, for both desktop and mobile devices.
The median of dwell time tended to differ greatly between desktop and mobile devices if the website was designed with only one type of device.
We also found that hour and month greatly affected the dwell time, but day of the week had little effect.
In particular, the change in dwell time and in the number of sessions (page views) was affected by the hour of day.
Moreover, we found that inside and click users tended to have longer dwell times than outside and non-click users.
This is likely because the level of interest affects the dwell time.
However, we can not find a relationship between dwell time and scroll depth.
Even if a user browsed to the bottom of the page, the user may not necessarily have read the entire page.

\bibliographystyle{IEEEtran}
\bibliography{IEEEabrv,references}

\begin{thebibliography}{1}
\providecommand{\url}[1]{#1}
\csname url@samestyle\endcsname
\providecommand{\newblock}{\relax}
\providecommand{\bibinfo}[2]{#2}
\providecommand{\BIBentrySTDinterwordspacing}{\spaceskip=0pt\relax}
\providecommand{\BIBentryALTinterwordstretchfactor}{4}
\providecommand{\BIBentryALTinterwordspacing}{\spaceskip=\fontdimen2\font plus
\BIBentryALTinterwordstretchfactor\fontdimen3\font minus
  \fontdimen4\font\relax}
\providecommand{\BIBforeignlanguage}[2]{{%
\expandafter\ifx\csname l@#1\endcsname\relax
\typeout{** WARNING: IEEEtran.bst: No hyphenation pattern has been}%
\typeout{** loaded for the language `#1'. Using the pattern for}%
\typeout{** the default language instead.}%
\else
\language=\csname l@#1\endcsname
\fi
#2}}
\providecommand{\BIBdecl}{\relax}
\BIBdecl

\bibitem{Agichtein2006}
E.~Agichtein, E.~Brill, and S.~Dumais, ``{Improving web search ranking by
  incorporating user behavior information},'' in \emph{Proceedings of the 29th
  annual international ACM SIGIR conference on Research and development in
  information retrieval}, 2006, p.~19.

\bibitem{Morita1994}
M.~Morita and Y.~Shinoda, ``{Information Filtering Based on User Behavior
  Analysis and Best Match Text Retrieval},'' in \emph{Proceedings of the 17th
  annual international ACM SIGIR conference on Research and development in
  information retrieval}, 1994, pp. 272--281.

\bibitem{Yi2014}
X.~Yi, L.~Hong, E.~Zhong, N.~N. Liu, and S.~Rajan, ``{Beyond clicks: dwell time
  for personalization},'' in \emph{Proceedings of the 8th ACM Conference on
  Recommender systems}, 2014, pp. 113--120.

\bibitem{Lalmas2015}
M.~Lalmas, J.~Lehmann, G.~Shaked, F.~Silvestri, and G.~Tolomei, ``{Promoting
  Positive Post-Click Experience for In-Stream Yahoo Gemini Users},'' in
  \emph{Proceedings of the 21th ACM SIGKDD International Conference on
  Knowledge Discovery and Data Mining}, 2015, pp. 1929--1938.

\bibitem{Zhou2016}
K.~Zhou, M.~Redi, A.~Haines, and M.~Lalmas, ``{Predicting Pre-click Quality for
  Native Advertisements},'' in \emph{Proceedings of the 25th International
  Conference on World Wide Web}, 2016, pp. 299--310.

\bibitem{Liu2010}
C.~Liu, R.~W. White, and S.~Dumais, ``{Understanding web browsing behaviors
  through Weibull analysis of dwell time},'' in \emph{Proceeding of the 33rd
  international ACM SIGIR conference on Research and development in information
  retrieval}, 2010, p. 379.

\bibitem{Lagun2016}
D.~Lagun and M.~Lalmas, ``{Understanding User Attention and Engagement in
  Online News Reading},'' in \emph{Proceedings of the Ninth ACM International
  Conference on Web Search and Data Mining}, 2016, pp. 113--122.

\bibitem{Song2013}
Y.~Song, H.~Ma, H.~Wang, and K.~Wang, ``{Exploring and Exploiting User Search
  Behavior on Mobile and Tablet Devices to Improve Search Relevance},'' in
  \emph{Proceedings of the 22nd International Conference on World Wide Web},
  2013, pp. 1201--1212.

\end{thebibliography}

\end{document}